\documentclass[journal,article,submit,oneauthor,pdftex]{mdpi} 

\firstpage{1} 
\pubvolume{xx}
\issuenum{1}
\articlenumber{5}
\pubyear{2019}
\copyrightyear{2019}
\history{Received: date; Accepted: date; Published: date}

\Title{Slim accretion disks: theory and observational consequences}

\Author{Bozena Czerny $^{1}$\orcidA{}}

\AuthorNames{Bozena Czerny}

\address{%
$^{1}$ \quad Center for Theoretical Physics, Polish Academy of Sciences, Al. Lotnikow 32/46, 02-668 Warsaw, Poland; bcz@cft.edu.pl}

\corres{Correspondence: bcz@cft.edu.pl}

\abstract{Slim accretion disks idea emerged over 30 years ago as an answer to several unsolved problems. Since that time there was a tremendous increase in the amount of observational data where this model applies. However, many critical issues on the theoretical side remain unsolved, as they are inherently difficult. This is the issue of the disk stability under the radiation pressure, the role of the magnetic field in the energy transfer inside the disk and the formation (or not) of a warm corona, and outflows. Thus the progress has to be done both through further developments of the model and through the careful comparison to the observational data.}

\begin{document}
\section{Introduction}

Slim accretion disk models were proposed in 1988 by \citet{abramowicz1988}. They describe an optically thick, geometrically not 
very thin quasi-Keplerian accretion flow onto a black hole in high Eddington ratio sources. In such a flow considerable part of the energy dissipated in the 
disk interior is carried radially with the flow instead of being reemited at the same radius, as in the standard \citet{ss1973} accretion disk model.

Observed sources are mostly ub-Eddington, but a fraction of active galactic nuclei and some galactic binaries during some stages are close to or above the Eddington ratio and
the slim disk model should apply there. Observations indeed support the presence of the optically thick disks in these sources. However, important 
issues related to the stability of these models and necessary modifications is still under debate. Slim disk of \citet{abramowicz1988} is based on 
so-called $\alpha$-viscosity assumption, while fully self-consistent model should predict the viscous torque. The 3-D numerical magneto-hydrodynamical 
simulations do not yet have all the necessary ingredients to show realistically how the flow onto the black hole proceeds, although there was a tremendous 
progress in this direction. Thus, at this moment, also semi-analytical models are also very useful tool in confronting the models with the 
observational data. One of the key issues is actually the presence of the heartbeat states in some astronomical sources and their relation (or not) with
the limit cycle behaviour predicted by slim disk model.

\section{Accretion disk models}

The key parameters in accretion process onto a black hole are the angular momentum of the inflowing material and the
accretion rate. Here we concentrate on high angular momentum flow which leads to formation of an accretion
disk, and the azimuthal velocity of such flow at a given radius is of the order of Keplerian, apart from the
innermost region of the flow close to ISCO.

The accretion rate is conveniently expressed using the dimensionless quantity, $\dot m$, the ratio of the accretion rate to the
Eddington accretion rate, $\dot M_{Edd}$. The formal definition of the last quantity is
\begin{equation}
\dot M_{Edd} = {L_{Edd} \over \eta c^2},
\label{eq:efi}
\end{equation}
where the Eddington luminosity is defined as usually (for a spherically symmetric setup, pure hydrogen plasma)
\begin{equation}
L_{Edd} = {4 \pi GM m_pc \over \sigma_T },
\end{equation}
$M$ is the black hole mass, $c$ is the light speed, $\sigma_T$ is Thomson cross-section for scattering in a fully
ionized material and $m_p$ is the proton mass, and $\eta$ is the dimensionless accretion efficiency. 

Here the subtle point is that the accretion flow efficiency of a slim disk itself depends on the accretion 
rate. In \citet{ss1973} model the accretion efficiency is 1/12, in full GR extension of this model 
\citep{novikov1973} it is uniquely given by the spin of the black hole, but in slim disks part of the energy
is trapped, and as the accretion rate rises, the fraction advected under the horizon rises, and the efficiency
of accretion flow drops. Therefore, some authors use $\eta=1.$ in Eq. \ref{eq:efi}, some use the 'typical value'
of 0.1, and finally some calculate it indeed from the model. In the discussion below I will generally use $\eta = 0.1$
as a reference value.

Currently used accretion disk models are thus divided predominantly according to the Eddington ratio of the flow, although the value of the angular momentum at the outer disk (more precisely, its relation to the local Keplerian angular momentum)  also plays a role. The physics of the accretion process is well described in a number of reviews \citep[e.g.][]{abramowicz2013}, and the best introduction in a form of a book is offered by \citet{frank2002}. Here I shortly summarize the basic forms of accretion flows. 

  \subsection{spherical accretion and low angular accretion flow}

Spherical accretion flow has negligible angular momentum, and the best example is the Bondi flow. Significant part of the flow proceeds supersonically. The flow can be optically thin, like in the case of accretion in weakly active galaxies, or it can be very optically thick in the case of dynamical collapse of a star. If some angular momentum is present, the material meets the centrifugal barrier. Accretion can proceed without, or with, a shock formation. If the corresponding circularization radius is much larger than the Innermost Stable Circular Orbit (ISCO), the rest of the flow has to be described by another models.

\subsection{optically thin ADAF flow}

If the Eddington ratio is low, well below a few percent, and the angular momentum is a fraction of the Keplerian angular momentum the flow is optically thin, the ion temperature is close to the virial temperature, and the electron temperature (much smaller than the ion temperature in the innermost part of the flow) is determined by the interaction with ions and direct dissipation mechanisms heating electrons. Such a flow is characteristic for Low Luminosity AGN, and for very low luminosity states in galactic sources. Significant fraction of the energy is then advected towards the black hole (hence Advection-Dominated Accretion Flow - ADAF). Since electron temperature there is high, the flow is the efficient source of the X-ray emission.

\subsection{standard optically thick geometrically disk.}

When the Eddington ratio is higher, the disk material cools down efficiently due to higher density, and the flow becomes optically thick, emitting locally as a black body to a good approximation. The angular momentum of such flow is locally Keplerian (\citealt{ss1973}, for full General Relativity version see \citealt{novikov1973}). The disk emission dominates the UV band (for AGN) and soft X-ray band (for galactic sources). The inner radius of the flow is well described by the position of ISCO. Heat advection is negligible in these models, and the local emissivity is by the local dissipation enforced by the transfer of the angular momentum. Accretion efficiency of the flow is set by the position of ISCO, i.e. by the black hole spin. Such models well apply to (some) soft states in galactic sources, and to many quasars in the optical/UV band.

\subsection{slim disks}

When the accretion rate becomes higher than the Eddington rate, the assumptions underlying standard model break down. Advection term becomes important, the inner radius of the disk moves from its ISCO position (mostly inward, but the effect depends on the viscosity), the departures from the Keplerian angular momentum due to the presence of strong radial pressure gradient become important. The local emission is no longer given by the local dissipation, and a fraction of energy is lost through the inflow under the black hole horizon. The accretion efficiency is lower than in the standard disk, and it decreases with the rise of the accretion rate. These models apply to super-Eddington quasars, Narrow Line Seyfert 1 galaxies, and also to some phases of gamma-ray bursts (but then the physics has to be modified to include neutrino cooling and nuclear processes), and to Tidal Disruption Events (TDE).

\subsection{transitions between the models}

Here slim disks and standard disks belong to a single branch of solutions. Full slim disk equations can be used to describe the standard disks. The additional terms, present in the slim disk, with automatically become relatively unimportant when the accretion rate is well below the Eddington. This is not the case for the transitions between optically thin and optically thick models. If the viscosity parameter is very high (close or above 1) formal bridge between ADAF and slim disk is found (skipping the standard gas-dominated disk stage), but for lower (more realistic) values the two branches are separate, as shown already in \citet{abramowicz1995}. Thus, the transition between the standard cold disk and the inner ADAF is still under debate, and it seem to require more physics like electron conduction to describe this process \citep{rozanska2000,liu2002}, although attempts of describing the transition just at the basis of heating/cooling change are also done \citep[e.g.][]{drew2017}. 

\section{Historical remarks}

Slim disk idea developed out of three earlier independent lines of studies which later convereged into a self-consistent
and practical model.

The oldest line of research focused on the proper description of the inner boundary condition of a standard Keplerian disk.
The classical model of Shakura \& Sunyaev \citep{ss1973} assumed that there is no torque acting at the inner edge of the
flow and concentrated on the description of the disk at larger radii, not even touching the issue how the flow actually
proceeds from the disk toward the black hole horizon. The issue was rised in a number of papers but without offering a
satisfactory solution \citep[e.g.][]{stoeger1976}. Lead by Bohdan Paczynski, we started the research in this direction at Copernicus
Astronomical Center, starting from the proper understanding that the flow must be actually the same as the outflow
through the inner Lagrange point L1 in binary starts. \citet{paczynski1981} used a simplified version of equations 
which did not allow yet to describe the transition of the flow from subsonic outside to supersonic flow below 
the innermost stable circular orbit, \citet{loska1982} defined
conditions of the transonic flow in the case of barytropic gas, and the complete description of the flow, with
advection term, radial radiative term, equation of state based on self-consistently determined gas to radiation
pressure ratio was presented by \citet{muchotrzeb1982}. The model described using the vertically-averaged disk model,
and the gravity field of the black hole was described in pseudo-Newtonian approximation \citep{paczynski1980}. Solution had to be determined numerically, and the code prepared by
\citet{muchotrzeb1982} was actually used by \citet{abramowicz1988} as it already has all the required elements, and it
already implied that with the rise of the accretion rate for a given black hole mass, the sonic point moves inward from ISCO
(Innermost Stable Circular Orbit) if the viscosity coefficient is small (for large viscosity behaviour,
see \citealt{muchotrzeb1983}).

The second line of research concentrated on geometrically thick accretion disks, known as Polish dognuts, which were seemingly required to
produce well collimated extended jets observed in a fraction of active galaxies.  This line of research was also vigorously
studied in Warsaw \citep[e.g.][]{jap1980}. An important contribution here to slim disk concept was to notice that at certain regime (above
the Eddington limit) the radiation in the (spherically symmetric) flow is trapped \citep{katz1977,begelman1978}. These models were truly two-dimensional, optically thick disks, and their construction was strongly based on the general idea of figures of equilibrium. For example, models created by \citet{paczynski1980} using the pseudo-Newtonian potential had an inner and also the outer edge, their radial extension was determined by the adopted radial distribution of the angular momentum, since the inner and the outer edge were located at the interception points with the Keplerian angular momentum distribution. Finding the equipotential surface passing through the inner and outer radius allowed to obtain immediately the full 2-D shape. The local radiation from the disk surface was calculated assuming that the local radiation flux, perpendicular to the disk surface is equal to the Eddington flux. Thus, overall emission corresponded to super-Eddington luminosity due to geometry. Global energy budget provided the corresponding accretion rate. However, local energy budget between heating and cooling is not considered in these models, and therefore the angular momentum distribution has to be assumed. In slim disks, developed later, this budget is considered, and therefore the angular momentum distribution is calculated self-consistently for an assumed accretion rate. These models are not used to interpret the data since they require these additional assumptions, but they are useful as initial setup for MHD time-dependent computations. An important contribution here to slim disk concept was also to notice that at certain regime (above
the Eddington limit) the radiation in the (spherically symmetric) flow is trapped \citep{katz1977,begelman1978}. It also showed
a geometrical similarity between an inner edge of geometrically thick disk and a star filling its Roche lobe.

The third line of research was devoted to the problem of thermal and viscous instability of a standard disk. These two
instabilities were discovered (soon after publication of \citealt{ss1973}) by \citet{prp1973} and \citet{le1974}, correspondingly
(for combined effect of the two instabilities, see \citealt{ss1976}).
This questioned the very existence of the accretion disks, otherwise attractive as explanation of the accretion phenomenon
in binary systems and in active galactic nuclei \citep{ss1973,lynden_bell1969}. Using unstable models seemed problematic.
The way out was first searched for in the form of modification of the viscosity law. The classical \citet{ss1973} model is based on
assumption that the viscour torque is proportional to the total pressure in the disk interior, i.e. gas plus radiation pressure, and
the instability appeared for relatively high accretion rate, when the radiation pressure starts to dominate. \citet{sakimoto1981}
proposed pure magnetic viscosity model and argued that in such case the viscous torque is proportional only to the gas pressure which
led to stable disks, of much higher surface density, with self-gravitational instabilit in the outer parts.  \citet{abramowicz1988}
introducing slim disk still adopted the \citet{ss1973} viscosity prescription but included the advection term which acts
(predominantly) as a cooling term. Thus, at very high accretion rates the disk becomes stable. This discovery brought back
the interest in accretion disks models as the models of the continuum emission in binary black holes and active galactic nuclei.

\section{Applicability and observational appearance}

The slim disk equation in their original formulation \citep{muchotrzeb1982,abramowicz1988} were using the pseudo-Newtonian gravitational potential of \citet{paczynski1980}, and they used the vertically-averaged disk structure. But they contained the key elements: the advection term, the radial pressure gradients, and the proper description of the inner boundary conditions. The presence of the radial pressure gradient allowed to determine, instead of just assume, the radial angular momentum distribution of the angular momentum in the disk. Specifically, the presence of these terms allow to determine where the radial velocity of the flow becomes supersonic, and it must become supersonic before it crosses the black hole horizon. As an inner boundary, the zero-torque condition is assumed, but not at ISCO, as in the standard disk, but at the black hole horizon. The flow is thus calculated all the way to the black hole horizon, and the location of the sonic point (usually below ISCO) results from the model and depends on the accretion rate and viscosity. Due to the presence of these additional radial terms, the model also describes the heat advected with the flow. As mentioned before, for accretion rates much lower than the Eddington ratio the model is in practice similar to the standard model, the advection term is almost unimportant, apart from the region where the flow changes into supersonic, but this transition happens very close to ISCO, and not much energy is dissipated there.

Further numerical developments cover formulation of the slim model in full GR, which gives the opportunity to study the dependence of the model on the black hole spin \citep{sadowski2011} since pseudo-Newtonian approach mimics well only the Schwarschild solution. In addition, \citet{sadowski2011} also introduced a description of the vertical structure of the disk by combining slim disk equations in the vertical plane with the disk vertical structure equations from Newtonian model of \citet{rozanska1999} which include the vertical radiative transfer, convection in the vertical direction, local description of the heating using the $\alpha$ parameters. Such 1+1 structure is the most advanced stationary slim model built up to date. However, many trends can be studied using much simpler models, like for example semi-analytical models like super-critical self-similar solutions \citep{wang1999}.

The role of advection in the disk rises with the rise of the accretion rate, $\dot m$, and some effects start to be
seen at $\dot m = 0.3$, and for $\dot m$ above 1 the effects are very important. The advection modifies the
shape of accretion disk spectrum. The long wavelength range is not strongly affected since this part of emission comes
from the outer parts where advection is not yet strong, but the innermost part of the disk emits less than in
classical model so the slim disk spectrum is redder than the corresponding standard model. Finally, the radiation
flux emitted close to ISCO in slim disk is again enhanced \citep[see Fig. 7 in][]{muchotrzeb1982} but this affects less
the total spectrum since only a relatively small fraction of energy is dissipated there. The effects were
studied in a number of papers, and models were applied to a whole range of sources from binary stars, through
intermediate black hole mass candidates to active galaxies \citet{szuszkiewicz1996,straub2014}. However, observationally the issue is not quite
settled, for example recent work on fitting broad band spectra of high Eddington ratio and lower Eddington ratio sources
did not show any clear differences, and both were well fitted by a standard disk model \citep{castello2017}.
The improvement in use of the slim disk over the standard disk was also not so significant in the case of
binary systems close to the Eddington luminosity \citep{straub2011}, although the model used there, developed
by \citet{sadowski2011}, was already an improvement of the original model of \citet{abramowicz1988}.

The slim disk model predicts that as the accretion rate rises, the luminosity rises relatively slowly, and finally
saturates at the value of about 10 times higher than the Eddington luminosity. The luminosity rise with accretion rate (see e.g. Figure~11 in \citealt{abramowicz2013}) above the Eddington rate can be well approximated as a logarithmic rise
\begin{equation}
L = 2 L_{Edd}(1 + ln(\dot m/5),    
\end{equation}
for a non-rotating black hole \citep{wang_shadow2014}.
    
AGN approaching this limit
were thus proposed to be used as standard candles in cosmology to measure the expansion rate of the Universe
\citep{wang_Edd_2014,marziani2019}.

Accretion disks at high accretion rate becomes geometrically thicker which additionally collimated the radiation
emitted by the innermost part of the disk and affect the illumination of the disk surrounding, including the
Broad Line Region \citep{wang_shadow2014}.

Determination of the black hole mass and Eddington ratio in large quasar samples are never very accurate, most of
the higher quality measurements from \citet{shen2011} catalog imply that most quasars concentrate around
$L/L_{Edd} \sim 0.1$ but the tail goes to or above the Eddington ratio \citep{panda2018}. Eddington or super-Eddington
accretion rates were claimed for a number of sources \citep[e.g.][]{collin2004,du2015,negrete2018,martinez2018}, and the results or the reverberation
measurements in super-Eddington AGN were different that similar results for lower Eddington ratio sources
\citep{du2016,du2018} supporting the view that indeed a fraction of the sources belongs to this category.

\section{Stability of classical slim accretion disks}

\label{sect:stability}

Altough the aim of formulating the slim disk model was to provide a stable solution for the high accretion rate flow, this
goal was not actually fully achieved. However, the most important step was made in unswering the question: what happens when
the disk becomes unstable due to radiation pressure? The answer was: at {\it sufficiently high} accretion rate indeed the flow
cools down by advection and the flow is stabilized. However, the global picture is more complex, in close analogy to the
ionization instability issue in the case of cataclysmic variables and X-ray novae.

The global behaviour of the accretion disk is governed by two trends: one describing the local behaviour of the disk structure 
changes with the rise of the accretion rate, and the second which shows how this behaviour scales with the distance from the black hole.

\begin{figure}[H]
\centering
\includegraphics{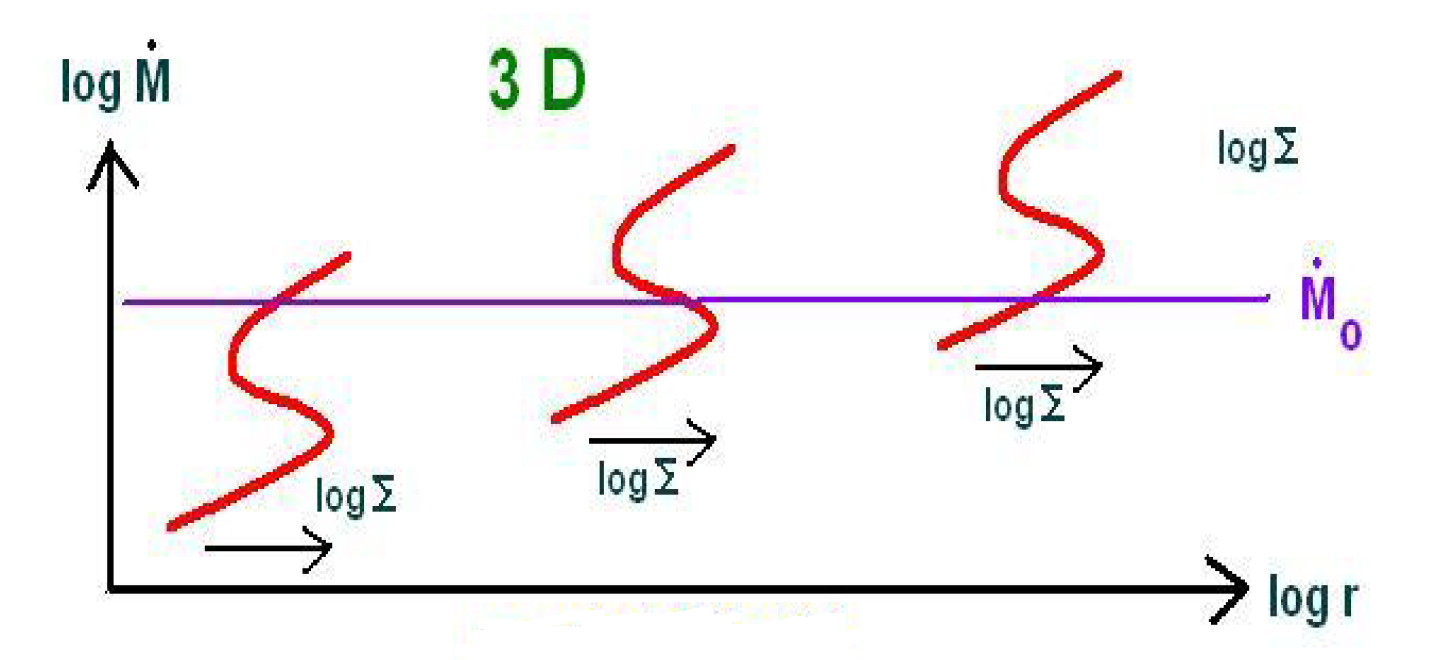}
\caption{Shematic illustration of the stability curve of an optically thick disk at few distance from the black hole. 
A single $\Sigma - \dot M$ picture shows the dependence of the disk surface density, $\Sigma$, on the accretion rate $\dot M$, 
the positive slope describes the stable branch (gas pressure-dominated lower branch, advection-dominated upper branch), the 
intermediate radiation-pressure dominated branch is unstable. The whole curve position shifts up with radius, so when we 
look at the location where the external assumed accretion rate crooses the plots we see (for high accretion rate) an inner, stable 
advection-dominated branch, and intermediate unstable zone and an outer stable gas dominated disk.}
\end{figure}

It we study locally the disk structure close to a black hole, for example at 10 $R_{Schw}$ (where $R_{Schw} = 2 G M/c^2$) as a function of the
accretion rate, at low accretion rate the disk surface density rises, when the pressure is dominated by the gass pressure. With further rise
of the accretion rate we enter the radiation-pressure dominated branch, and the disk surface density now decreases with the rise of the accretion rate.

This solution is unstable, since the heating rises more rapidly than the disk cooling with a small rise in the disk temperature. Finally, when we
approach the Eddington limit the advection sets in, and a new stable advection-dominated branch developes. Now again the surface density in the disk 
rises with the increase of the accretion rate. Thus, for high accretion rates the disk becomes stable. This is quantitatively 
ilustrated in Figure 2 of \citet{abramowicz1988}, and schematically shown in Figure 1 above, first insert. Since the $\dot M - \Sigma $
plots shows so nicely the stability or instability of a partucular disk we call these plots stability curves.

\citet{abramowicz1988} plots the stability curves only in a very narrow range of radii. The upper branch of this curve (stable branch) corresponds to a slim disk solution while the other two branches are predicted by the classical disk model. The intermediate branch is unstable. In \cite{ss1973} this branch would extend till arbitrary accretion rate but the new terms present in the slim disk model (negligible at low accretion rates) forced the unstable branch to bent. However, if we go to radii of order of a few hundred 
$R_{Schw}$, the overall shape of the curve does not change much, but the 'location' of the curve changed dramatically. This is 
schematically illustrated in Figure 1. The dominating effect is the shift up, so if we consider an accretion rate of a few (in dimensionless units, 
introduced above), at a distance of hundreds of $R_{Schw}$ such accretion rate would imply that the disk is at the unstable radiation-pressure 
dominated branch, and advection is still unimportant, while if he go further from the black hole, for the same accretion rate the disk 
is still dominated by the gas pressure. This happens since in general, in the classical solution of the \citet{ss1973} the ratio of the radiation to 
gas pressure goes down with the radius for a fixed accretion rate. The radii when these transitions take place depend on the mass of the central body, 
and in general the accretion disks in active galaxies are much more radiation-pressure dominated than in the case of accretion disks at binary black holes.
The stabilization due to advection happens in both at roughly the same accretion rate but the stable and unstable zones are much more extended in active 
galaxies, as illustrated in \citet{janiuk2011}.

Therefore, unless the outer radius of the disk is very small, slim disks always have intermediate unstable zone. Thus the models are not stationary. Still, due to the new
advection term it was later possible to calculate the time evolution of the disk in the same way as it was done for cataclysmic variables 
\citep{meyer1981,smak1984}, where the
stable upper and lower branch were provided by neutral and fully ionized disk states, complementing the intermediate unstable 
branch, corresponding to parially ionized gas. Only for a slim disk we have an upper advection-dominated branch and lower, gas pressure dominated branch.

If, at a given radius, the mean accretion rate locates the disk on the unstable (radiation-pressure dominated) branch, the disk structure performes the time evolution, with epizodes of
outbursts and low states. During outburst the source is on the upper branch, the temperature and radiation flux are high, the accretion rate is high but the ring 
looses effectively the material, as the mass inflow 
towards the black hole is not fully compensated by the inflow from outer radii. The disk surface density systematically decreases during this phase. At low state, 
the temperature and the radiation flux are low, the material effectively accumulates in the ring, and the surface density rises slowly in a viscous timescale.

The limit cycle must be happening ati-clock wise, on the stability curve, as the arguments above imply. The transition from the upper to lower branch takes place when the
disk cannot achieve a solution in thermal balance on the upper branch with further reduction of the surface density, and the transition takes place in the thermal 
timescale. Roughly the same happens when the ring accumulated so much mass that further increase of the survade density leand to lost thermal balance and the disk rapidly 
expands reaching the upper branch.

Of course, rings are coupled, so it is actually necessary to calculate the evolution of the whole disk but it looks qualitatively similar: large parts of the disk 
alternate between the upper and the lower stable branch. The inner, formally stable part of a slim disk is also affected since the outer - unstable - part of the disk
alternates the supply of the material to the innermost part. So whole inner disk is subject to strong variability.

The models of the time-dependent evolution has been calculated by a number of autors \citep{nayakshin2000,janiuk2000,janiuk2002,janiuk2007,czerny2009,grzedzielski2017a,grzedzielski2017b}, generally with the aim to compare to some observational data for 
objects showing outbursts. Thus, to fit the outburst amplitude, some modification of the model were usually introduced since in general the outbursts 
calculated without any modification had too high amplitudes. We will return to this issue later on. In general, these simulations 
showed regular or semi-regular changes of the predicted luminosity up to a few orders of magnitude. Computations were done in 1-D approximation but globally,
with calculated total duration of several of the {\sl viscous} timescales at the outer, stable part of accretion disk but they still resolved the time evolution
of the inner disk part in the thermal timescale.

Here an important point was raised by \citet{gulu2007}. The equations of the slim disks were always
solved still in geometrically thin disk approximation, i.e. with expansion of the gravitational potential against $z/r$, where
$z$ is the distance from the equatorial plane. This might not be correct for high Eddington ratio sources, since the prediction of the standard disk is that the disk thickness in the radiation pressure dominated part rises linearly with the accretion rate \citep{ss1973}. If the local disk thickness become larger than the radius, the gravity force in the verical direction would start to decrease with further rise of accretion rate. This in turn might lead to a strong outflow.
However, \citet{lasota2016} argued that this never happens since the advection term
will prevent slim disk from becoming thick at arbitrary high accretion rate. Thus, the standard computations of the slim disk 
evolution are self-consistent.

\section{Observational support of outbursts due to radiation pressure}

The issue is still under debate whether some regular outbursts observed in accreting sources are actually signatures of the radiation pressure 
instability in slim disks. In the class of binary black holes, two sources display well defined periodic outbursts (now refered to as heartbeat mode):
GRS 1915+105 and IGR J17091-3624. The outbursts last of order of seconds to hundreds of seconds. Several properties of the outbursts are well modeled by
the radiation pressure instability, transitions between the states take place anti-clockwise \citep{janiuk2002}, outbursts are longer when the mean 
accretion rate is higher \citep{janiuk2000}. Also the time delay between the hard X-rays and the sift X-rays are in agreemnet with the data if a model of 
an unstable disk with an accreting corona is used for GRS 1915+105 \citep{janiuk2005}. On the other hand, other sources do not show such clear states, 
even at high Eddington ratios, which can be either interpreted as an argument 
against the very existence of the radiation pressure instability, or might imply that in other sources the effect of winds and/or magnetic fieled changes
the regular outbursts into semi-regular periods of enhanced variability \citet{janiuk2011}.

Active galaxies are much larger in size, so the timescales of the same phenomenon are correspondingly longer, and the predicted durations of outburst 
caused by radiation pressure are of order of thousands of years, and they are not accesible to direct observations. However, statistical comparison of 
the duration of active phave in short-lived radio loud AGN is generally consistent with expectations of the slim disk instability \citep{czerny2009}.
 \citet{wu2016} argued that this is a universal phenomenon seen across the whole black hole mass scale, from binary systems through 
intermediate black hole outbursts (source HLX-1) to active galaxies.

The comparison of the simple parametric models with the data is one of the ways to establish whether this instability is present. On the other hand, models
are based on assumption of the viscosity law, and usually some modifications are required to decrease the outburst amplitude, like the use of the $\alpha \sqrt{P_{gas}P_{tot}}$  or 
even more general models \citep[e.g.][]{szuszkiewicz1990,grzedzielski2017a}, or wind coupled to the local accretion rate \citep[e.g.][]{janiuk2015}.

\section{MHD simulations of radiation pressure dominated disks}

\citet{ohsuga2005} were the first to calculate the model of the superctical accretion flow. The flow settled
down to a quasi-stationary state. However, these computations were still done in radiative hydrodynamical (RHD) mode,
so used the viscosity prescription basically as in standard disk. Also the radial range considered
was too small to see any issues related to the global stability discussed above (input material had circularization radius at 100 $r_g$).

What is more important, since the year 1981 we are convinced
that the viscosity is generated as a result of magnetorotational instability (MRI; \citealt{balbus1991}). Therefore, the ad-hoc assumption of

the $\alpha$-viscosity law is not needed (in principle) any more. The
real progress and answer to the question about the global stability of the super-Eddington flow should come
from radiative MHD computations. With this insight, the arbitrary assumption of the viscous torque scaling is not needed, as
it should come automatically and self-consistently from the computations.

However, in reality the issue is not as simple as outlined above. If MRI instability has to be followed, the computations of the 
disk structure must be done in 3-D (even 2-D is not a good approximation for proper desciprion of the dynamo action). The 
computational timestep must be also shorter than in 1-D computations discussed in Section \ref{sect:stability}. So instead of a single PC,
computations are to be run at computer clusters, and they have difficulties in covering the whole disk for many viscous timescales.

Most 3-D MHD simulations are actually performed in a shearing box approximation, when only a very narrow radial zone is included, and periodic 
boundaries are assumed. This apparently affects the results. First results suggested that radiation pressure instability does not develop in 
real radiation pressure dominated disks \citep{turner2004,hirose2009} but subsequent sumulations using larger grid domain and more advanced radiative transfer approximation
were able to see the thermal runaway after a few thermal timescales \citep{Jiang2013}. \citet{Jiang_bump_2016} concluded that if the opacity in the computations include the atomic transitions of iron ions instead of just Thomson cross-section, 
the disk is stabilized, but again they used a very narrow ring in their computation and the global stabilization of the disk by the use of 
more complex opacity is unlikely \citep{grzedzielski2017b}.

\section{Hot coronae,warm coronae,outflows}

However, the most important issue of the slim disk model, and its limitations are related to topics we still do not understand. 
However, observations show us clearly that our knowledge is incomplete.
Slim disks, like the classical disks, do not provide explanation for the jet formation, hard X-ray emission,
soft X-ray excess and winds. This is because the disk is basically assumed to be dominated by the thermal pressure, and MRI can change it into
outer optically thin layers which do not contribute much to the total spectrum. On the other hand, spectra and variability implies that
that these phenomena take place in the inner part of the disk. So something must be missed in the model.

Adding these elements to the model in a parametric way modifies the behaviour of the disk itself, including the disk stability. Winds can reduce the outburst magnitude, 
as mentioned already in Section \ref{sect:stability} but magnetic field providing the energy transport deeply inside the disk can completely stabilize the 
disk. Such
models in parametric form were proposed by a number of authors, particularly in the context of the corona formation and hot flow/cold flow transitions \citep[e.g.][]{svensson1994,zycki1995,chakrabarti1995,rozanska1999,czerny2003,rajesh2010}. 

More
advanced, but still semi-analytical models of this phenomena were recently developed, for example by \citet{begelman2015,begelman2017,dexter2019,gronkiewicz2019}.  Some 3-D simulations of such flows were also 
performed \citep[e.g.][]{fragile2017} but they required ad hoc presence of the large scale magnetic fields which did not result from
the simulations. 

More complex time depended behaviour of the disk interior can also stabilize the disk. \citet{janiuk2012} considered a role of the stochastic fluctuations of the disk which are naturally expected due to MRI-based viscosity mechanism. They showed that large enough amplitude of such turbulence quenches the radiation pressure instability. This model is semi-analytical, and in principle, the described behaviour should be reproduced in advanced MHD computations without any {\it ad hoc} assumptions. But current MHD results are not yet capable of reproducing many other phenomena like power spectra, frequency-dependent time lags, which still have to be modeled in an alternative way \citep[e.g.][]{ahmad2018}.

Interesting possibility is related to the idea of Magneticall Arrested Disks (MAD) which could lead to modification of the flow close to the central object due to accumulated strength of the large-scale field lines.
Such phenomenon has been considered by a number of authors \citep[][]{narayan2003,mondal2019}, and it could be responsible for radio loud/radio quiet dychotomy in active galaxies, for the phenomenon of Changing Look (CL) AGN or
for the appearance of Ultra-Luminous X-ray sources. However, simulations of MAD were generally done in the context of optically thin ADAF-type flow, and not in the case of slim disks, and we do not know if indeed the same mechanism
can work also for those disks.

\section{Physics missing}

We may still be missing some key elements in the global 3-D MHD simulations. We actually do not know whether it is realistic to expect that a basic MHD
run of the accretion process will explain all the complexity observed in accreting black holes. However, this is certainly a goal, and at least we should try to test, if we have enough physics in
the equations to expect that. Some elements are still clearly missing.

For example, most recent 3-D simulations performed so far for sub-Eddington accretion rates by \citet{Jiang2019} include new elements: radiation viscosity in the optically thin corona region. This allows them 
to see the formation of the warm/hot corona, with temperature above $10^8 K$ in the innermost part of the disk, and the corona becomes more compact for rising accretion rate.

\citet{ghoreyshi2018} propose to add the radial viscous force. \citet{mckinney2017} show that double Compton and cyclo-synchrotron cooling is also important and affects the conclusions about the 
temperature in the innermost part of a slim disk. Convection is certainly present in the accretion disks \citep[e.g.][]{rozanska1999,sadowski2009}. This should be present automatically in global
3-D simulations but not in shearing-box approximation.
Several processes are well known, but usually not included in the modelling due to their complexity, like electron conduction \citep[e.g.][]{liu2017}, magnetic field reconnection \citep[e.g.][]{dalpino2010}.

We may also need to return to the issue of the anomalous viscosity in the disks (i.e. $\alpha$-viscosity). It is nicely summarized by \citet{Hawley1998}. They stress that the standard dynamical voscosity
in the flow, $\nu$, when forced to provide the requested $\alpha$-viscosity has implications for the mean free path and the turbulent velocity of the medium:
\begin{equation}
\nu = \rho v_t L = \alpha \rho v_s H,
\end{equation}
where $\rho$ is the local density, $v_t$ is the turbulent motion velocity, $v_s$ is the sound speed, and $L$ is the mean free path of a particle, and $h$ is the disk thickness.

Since in \citet{ss1973} model, $\alpha$ is of order of 0.1,
this means that we usually assume a mean free path of the partcle of order of 0.1 H, and the turbulent motion velocity equal to that of the sound speed. This is extreme, and this is the reason
why the viscosity
operating in accretion disks was named 'anomalous'. MRI solved this problem, we do not need 'anomalous' viscosity any more. However, we should note, that
if hot two-temperature corona forms above an accretion disk, in such a corona mean free path of ions is actually quite large, and sound speed is latge there as well, so in the hot corona
standard dynamical viscosity should effectively provide an efficient accretion flow. This is not yet included in any 3-D simulations to my knowledge.


\section{Conclusions}

In this review I shortly presented the history of the slim disk model, its applicability range, and most of all, its limitations. 
The description of the accretion flow close to or above the Eddington limit is still quite uncertain. Further, more advanced MHD 
simulations are needed, but they have to be supplemented with simpler, easy to calculate 1-D models, which can be easily compared
to the broad band observational data for galactic binary systems and active galaxies. Modelling time-dependence in observed sources 
is particularly valuable since stationary spectra are frequently rather degenerate with respect to the model properties. We should 
keep in mind that spectral properties of a stationary, optically thick, geometrically thin Keplerian disk of \citep{ss1973} do not 
depend on the disk interior (including the viscosity law) while the time-dependent evolution reflect the internal structure. Slim disks
should also be tested not only against the spectral properties but also through their time variability.Parametric approach to modelling the data can also help us to know how frequently real objects are in this regime, so developments of models with slim disk option included \citep{kubota2019} and prepared to bo fitted to the data is very valuable.

\vspace{6pt} 




\funding{This research waspartially supported by the National Science Center, Poland, grant No. 2017/26/A/ST9/00756 (Maestro 9).}

\acknowledgments{We are grateful to Agnieszka Janiuk and Jina-Min Wang for very helpful recent discussions of the topic of slim disks.}

\conflictsofinterest{The author declares no conflict of interest.} 




\externalbibliography{yes}
\bibliography{slim}



\end{document}